\documentclass[a4paper,11pt]{article}
\usepackage{pos}

\DeclareMathOperator{\tr}{tr}

\title{Lattice techniques to investigate the strong $CP$ problem: lessons from a toy model}

\author*[a]{David Albandea}
\author[a]{Guilherme Catumba}
\author[a]{Alberto Ramos}

\affiliation[a]{Instituto de Física Corpuscular (CSIC -- University of
Valencia), Parque Científico, C/Catedrático José Beltrán, 2, 46980, Paterna,
Valencia, Spain}

\emailAdd{david.albandea@ific.uv.es}

\abstract{Recent studies have claimed that the strong $CP$ problem does not occur
in QCD, proposing a new order of limits in volume and topological sectors when
studying observables on the lattice. We study the effect of the
topological term on a simple quantum mechanical rotor that allows a lattice
description. We particularly focus on recent proposals to face the challenging
problems that this study poses in lattice QCD and that are also present in the
quantum rotor, such as topology freezing and the sign problem.}

\FullConference{The 41st International Symposium on Lattice Field Theory (LATTICE2024)\\
 28 July - 3 August 2024\\
Liverpool, UK\\}

\begin{document}
\maketitle

\section{Introduction}

The QCD Lagrangian admits an additional renormalizable and gauge invariant term
known as the $\theta$ term,
\begin{align}
\delta \mathcal{L}_{\theta} = \theta q(x) \equiv \frac{\theta}{16\pi^2} \tr F_{\mu\nu} \tilde{F}^{\mu\nu},
\end{align}
where $\theta\in [0, 2\pi)$, is
a free parameter of the model on which physical observables can potentially
depend. This term would violate $CP$ symmetry, but from experimental
measurements of electric dipole moment of the neutron we know that $\theta$ must be
extremely
small,
$\left| \theta \right| \lesssim 10^{-10}$~\cite{Abel:2020pzs,RevModPhys.91.015001}.
The puzzle why it happens to be so small is known as the strong $CP$ problem.

There have been many proposed solutions in the literature to explain the
smallness of the $\theta$ angle, such as a Peccei-Quinn symmetry~\cite{PhysRevD.16.1791} and Nelson-barr
type models~\cite{Nelson:1983zb,PhysRevLett.53.329}. However, the strong $CP$ problem is far from being settled, and there
have been repeated debates about whether the $\theta$ angle affects physics at all, regardless of
its value.  Particularly, a recent proposal
argues~\cite{ai_absence_2021,ai_consequences_2021} that expectation values in
infinite volume can be obtained from expectation values at
finite volume as
\begin{align}
  \label{eq:intro:new-proposal-limit}
\langle O \rangle = \lim_{N\to\infty} \lim_{V\to\infty} \sum_{\left| Q \right| <
N}^{} \langle O \rangle_{Q,V} \;p_{V}(Q),
\end{align}
where $O$ is an observable, $\langle \cdots \rangle_{Q,V}$ denotes expectation value at fixed
topological sector $Q$ and finite volume $V$, $p_{V}(Q)$ is the distribution
of topological charge at finite volume, and the volume is taken to infinity before the contributions from
all topological sectors are summed. The consequence of such an order of
limits, opposite to the conventional one, would be the absence
of $\theta$-dependence from observables, implying that there is no strong $CP$
problem.

The $\theta$-dependence is difficult to study in lattice QCD simulations because
of several computational challenges, such as the sign problem~\cite{deForcrand:2009zkb,Gattringer:2016kco}  and
topology freezing~\cite{Alles:1996vn,DelDebbio:2002xa,DelDebbio:2004xh,Schaefer:2010hu}. However, the claims presented in
Refs.~~\cite{ai_absence_2021,ai_consequences_2021} should also hold in simpler
models. The present work is based on Ref.~\cite{Albandea:2024fui} and is organized as follows: in Sec.~\ref{sec:qr}, we study the order of limits of Eq.~(\ref{eq:intro:new-proposal-limit}) in the
one-dimensional quantum rotor, showing that it disagrees with the conventional one for the particular case of the topological susceptibility; in Secs.~\ref{sec:top-freezing} and \ref{sec:sign-prob}, we introduce two recently proposed algorithms to overcome both topology freezing and the sign problem, respectively; finally, in Sec.~\ref{sec:results}, we study the continuum limit of the topological susceptibility and the $\theta$-dependence of the ground state of the spectrum of the quantum rotor with lattice simulations.

\section{The quantum rotor}
\label{sec:qr}

\begin{figure}[!t]
\centering
\includegraphics[width=\textwidth]{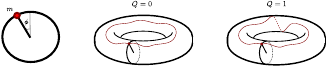}
\caption[Quantum Rotor topology]
{
(Left) Representation of the quantum rotor. (Middle and right) Possible trajectories of the
quantum rotor along a torus in the path integral representation, with
topological charges $Q=0$ and $Q=1$.
}
\label{fig:intro:rotor1}
\end{figure}

The quantum rotor is the simplest theory with topology and the presence of
a $\theta$ term: it is a free particle of mass $m$ on a ring, with
Hamiltonian
\begin{align}
H = - \frac{1}{2I} \left( \partial_{\phi} - \frac{\theta}{2\pi} \right)^2,
\end{align}
where $\phi$ is the angle representing the position of the particle in the ring
(see Fig.~\ref{fig:intro:rotor1} [left]),
$I = mR^2$ is its moment of inertia, and $\theta \in [0, 2\pi)$ is a free
parameter, analogous to the $\theta$ angle of QCD.

The system can also be formulated as a path integral at finite
temperature $\beta$ and Euclidean volume $T = 1 / \beta$ via the partition function~\cite{Fjeldso:1987wi,bietenholz_perfect_1997}
\begin{align}
  \label{eq:partition-function}
Z_{T}(\theta) = \int \mathcal{D}\phi\; e^{-S[\phi] + i\theta Q[\phi]},
\end{align}
where action and topological charge in the continuum read
\begin{align}
  \label{eq:intro:qr-cont-action}
S[\phi] =  \frac{I}{2} \int_{0}^{T} dt\;\dot{\phi}(t)^2, \quad Q[\phi] = \frac{1}{2\pi} \int_{0}^{T} dt \; \dot{\phi}(t) \in \mathbb{Z},
\end{align}
where $-\pi < \phi(t)\le \pi$ and periodic boundary conditions are imposed,
i.e. $\phi(T) = \phi(0) + 2\pi n$ with $n \in \mathbb{Z}$. Additionally, topology can be easily visualized in this model, as the periodic
boundary conditions make the trajectory of the particle live on the surface of
a torus (see Fig.~\ref{fig:intro:rotor1} [right]), where each trajectory can be
classified with an integer $Q$: if the trajectory of a configuration does not
wind around the torus, it has topological charge $Q=0$; if it winds once
around the torus, it has $Q=1$.

This model can be trivially solved using the quantum mechanical
formalism, and the energy levels of
the system read
\begin{equation}
  \label{eq:spectrum continuum}
  E_n = \frac{1}{2I} \left( n - \frac{\theta}{2\pi} \right)^2, \quad n\in\mathbb
  Z\,,
\end{equation}
from which one can build the thermal partition
function
$Z_{T}(\theta) = \sum_{n\in\mathbb Z} e^{- T E_{n} } = \sum_{n\in\mathbb Z} e^{- \frac{T}{2I}\left( n- \frac{\theta}{2\pi} \right)^2 }$
and obtain the probability distribution of each topological sector,
\begin{align}
p_{T}(Q) = \frac{1}{Z_{T}(\theta=0)} \frac{1}{2\pi} \int_{-\pi}^{\pi} d\theta\; Z_{T}(\theta) e^{-i\theta Q} = \frac{1}{Z_{T}(\theta=0)} \sqrt{\frac{2I\pi}{T}}\exp \left( -\frac{2I\pi^2}{T}Q^2 \right).
\end{align}
With these results one can test the order of limits of
Eq.~(\ref{eq:intro:new-proposal-limit}). Particularly, the topological
susceptibility in a finite volume $V$ can be computed as
\begin{align}
\chi_{t,V} = \frac{\langle Q^2 \rangle_{V}}{V},
\end{align}
and studying the infinite volume limit as in
Eq.~(\ref{eq:intro:new-proposal-limit}) one finds
\begin{align}
  \chi_t = \lim_{N\to\infty} \lim_{T\to\infty} \frac{1}{T} \sum_{\left| Q \right|<N} Q^2 \,p(Q)\,
  = \lim_{N\to\infty} \lim_{T\to\infty} \frac{1}{T}\frac{\sum_{|Q|\leq N} Q^{2}
  \exp( -\frac{2\pi^{2}I}{T}Q^{2})}{\sum_{|Q|\leq N} \exp(
-\frac{2\pi^{2}I}{T}Q^{2})},
    \label{eq:sum sectors}
\end{align}
which is trivially zero. Alternatively, one can obtain the topological
susceptibility in the zero temperature limit ($T\to\infty$) directly from the
energy spectrum in Eq.~(\ref{eq:spectrum continuum}), which
reads
\begin{align}
  \label{eq:intro:qrotor-top-susc-E0}
\chi_{t} \equiv \frac{d ^2 E_{0}(\theta)}{d \theta^2}\Big|_{\theta=0} = \frac{1}{4\pi^2 I}.
\end{align}
This result can also be obtained with the conventional order of
limits~\cite{Albandea:2024fui}, and contradicts the result coming from the
proposed order of limits in Eq.~(\ref{eq:sum sectors}). In the following, we
will validate the conventional order of limits using lattice simulations, for
which we will need to deal with topology freezing and the sign problem.

\begin{figure}[!t]
\centering
\includegraphics[width=0.7\textwidth]{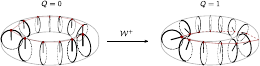}
\caption[Quantum Rotor topology]
{
Winding transformation on a quantum rotor configuration on the lattice.
}
\label{fig:intro:rotor-winding}
\end{figure}

\section{Topology freezing and the winding HMC algorithm}
\label{sec:top-freezing}

Standard algorithms for lattice QCD, such as the Hybrid Monte Carlo (HMC)
algorithm~\cite{DUANE1987216}, are well-known to suffer from topology freezing: near the
continuum limit, continuous update algorithms get trapped within a topological
sector, thus failing to sample the full configuration space and leading to
exponentially increasing autocorrelation times as the continuum limit is
approached in a finite volume.

The topology freezing problem is also present in simple models
such as the quantum rotor, making them useful testbeds for new algorithms aiming
to tackle this problem. The spacetime discretization of the quantum rotor
consists of $\hat{T} = T /a$ angle variables $\phi_{t}$
for $t\in \{ 0, \dots, \hat{T}-1 \}$ separated by a lattice spacing $a$.
Here we will work with the so-called standard discretization of the action and
topological charge,
\begin{equation}
    \label{eq:st-s-q}
  S_{\rm st}[\phi] = \frac{\hat I}{2}\sum_{t=0}^{\hat{T}-1}
  (1-\cos(\phi_ {t+1}-\phi_t))\,, \quad   Q_{\rm st}[\phi] = \frac{1}{2\pi} \sum_{t=0}^{\hat{T}-1}
  \sin(\phi_ {t+1}-\phi_t),
\end{equation}
where $\hat{I} = I / a$, as well as with the classical perfect
discretization,\footnote{Note that the classical perfect topological
charge, $Q_{\text{cp}}$, has a geometrical definition and is exactly an
integer.}
\begin{equation}
    \label{eq:cp-s-q}
  S_{\rm cp}[\phi] = \frac{\hat I}{2}\sum_{t=0}^{\hat{T}-1}
  ((\phi_ {t+1}-\phi_t)\bmod 2\pi)^2, \quad    Q_{\rm cp}[\phi] = \frac{1}{2\pi} \sum_{t=0}^{\hat{T}-1}
  ((\phi_ {t+1}-\phi_t)\bmod 2\pi) \in \mathbb{Z}.
\end{equation}
Note that both discretizations lead to the same continuum limit of Eq.~(\ref{eq:intro:qr-cont-action}) when taking $a \to 0$
along a line of constant physics with $T / I = \hat{T} / \hat{I} = \text{constant}$.

The simplest idea to build an algorithm that can sample topology efficiently is
to find transformations between topological sectors. Such a topology-changing
transformation can be easily built for the quantum rotor, for which we define the winding
transformation by
\begin{align}
\mathcal{W}^{\pm}:\; \phi_{t} \to \phi_{t}^{\mathcal{W}^{\pm}} = \phi_{t} \pm 2\pi t / \hat{T}, \quad \mathcal{W}^{+} \mathcal{W}^{-} = I,
\end{align}
where the $+$ (winding) or $-$ (antiwinding) is common to all $t\in[0,\hat{T}-1]$. As shown in Fig.~\ref{fig:intro:rotor-winding}, the transformation gradually
winds the trajectory of a configuration around itself exactly once, such
that $Q_{\text{cp}}(\phi^{\mathcal{W}^{\pm}}) = Q_{\text{cp}}(\phi) \pm 1$. This
transformation can be easily embedded into a Metropolis algorithm with
acceptance
\begin{align}
p_{\text{acc}}(\phi^{\mathcal{W^{\pm}}} \mid \phi) = \min \left\{ 1 , e ^{- S[\phi^{\mathcal{W}^{\pm}}] + S[\phi]} \right\},
\end{align}
and we denote the combination of this Metropolis step with the HMC algorithm as the
winding HMC (wHMC) algorithm~\cite{Albandea:2021lvl}, which restablishes
ergodicity within the full configuration space if the acceptance is significant.

\begin{figure}[t!]
	\centering
\includegraphics[width=0.6\textwidth]{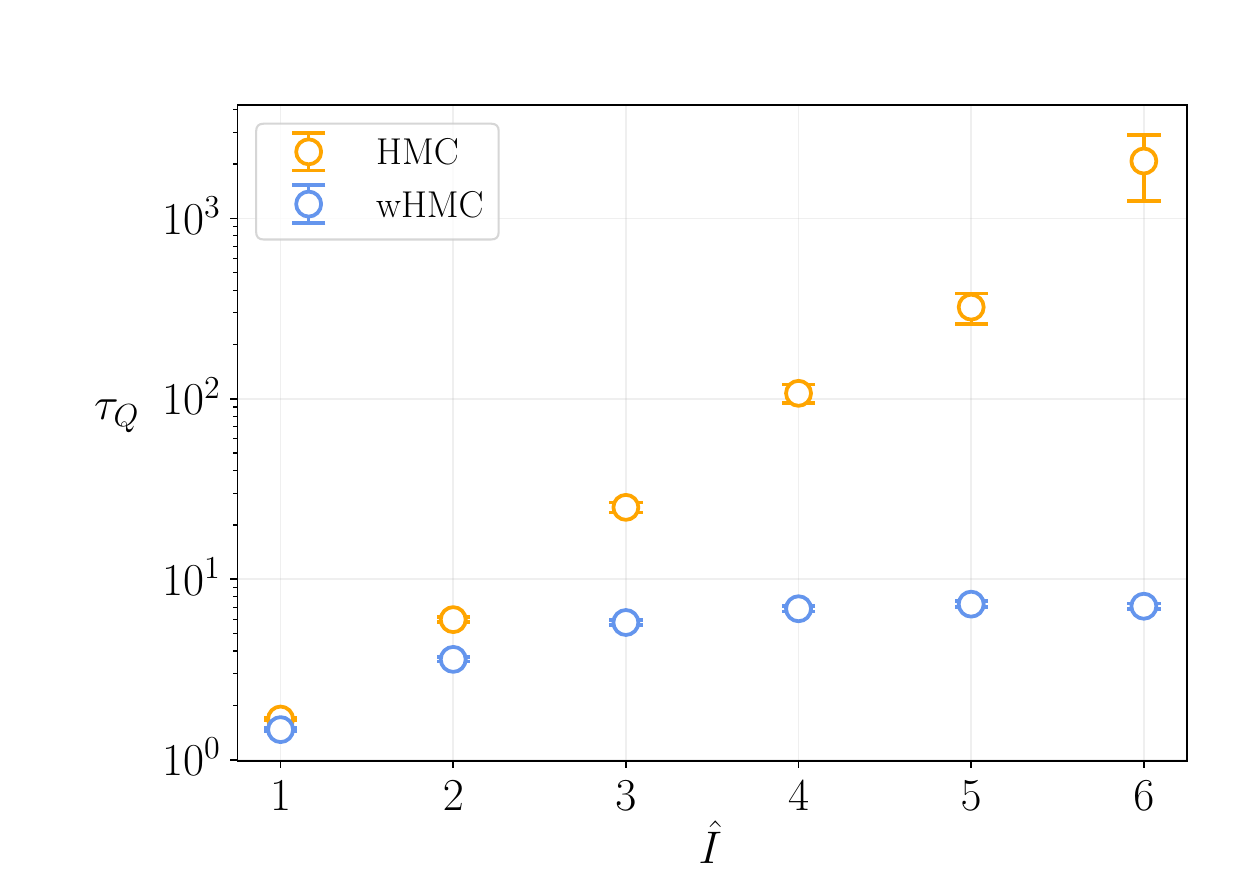}
\caption{Autocorrelation time of $Q$ as a function of $\hat{I}$ for wHMC (blue)
  and HMC (orange), keeping $\hat{T} / \hat{I} = 100$, with the standard
  discretization of the action.
  } \label{fig:windings:qrotor-tauiscaling-Q}
\end{figure}

In Fig.~\ref{fig:windings:qrotor-tauiscaling-Q} we show the scaling of
autocorrelation times of the topological charge, $\tau_Q$, from simulations with the HMC
and wHMC algorithms. While the autocorrelations of HMC scale exponentially, as
expected, the ones of wHMC eventually saturate, thus solving
the topology freezing problem in this model.

\section{The sign problem and truncated polynomials}
\label{sec:sign-prob}

\begin{figure*}[!t]
  \centering
 \includegraphics[width=0.49\textwidth]{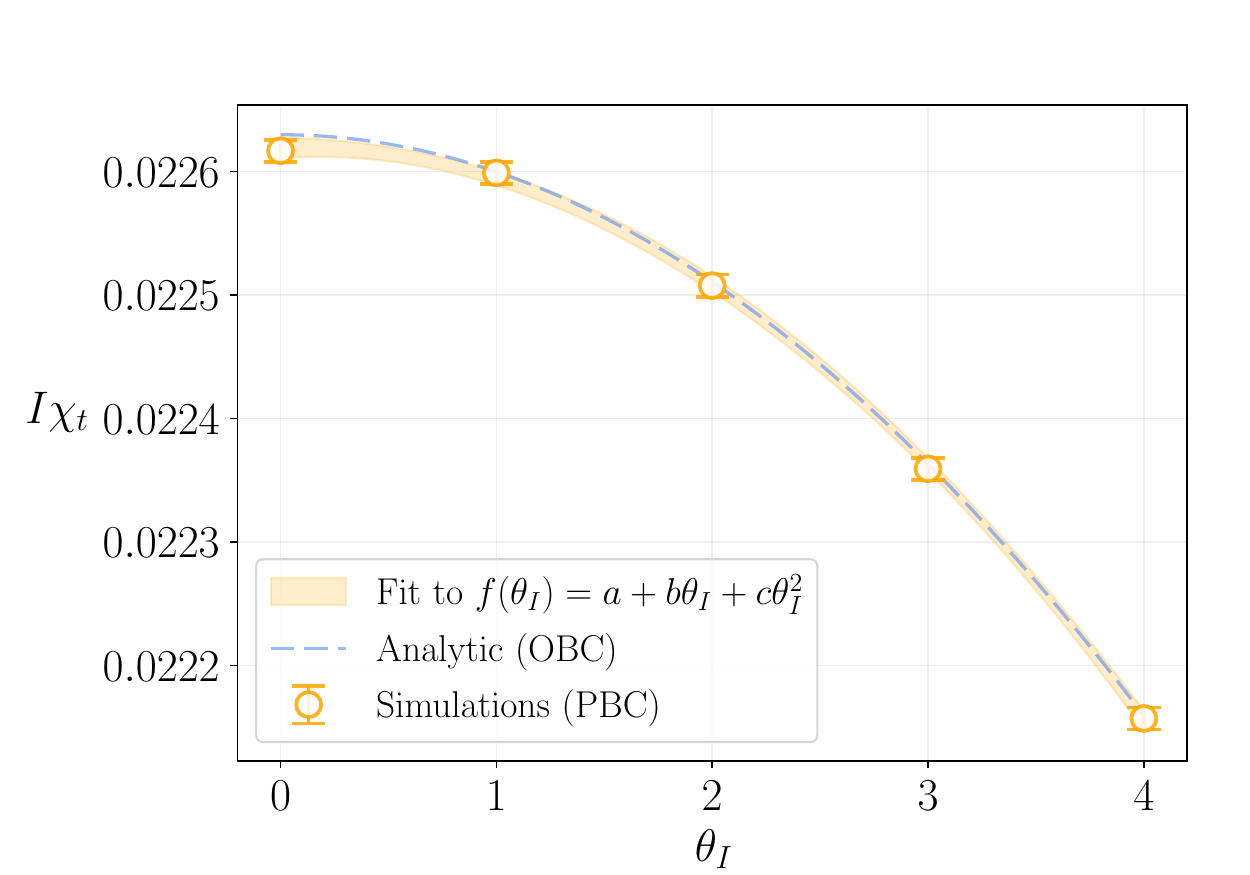}
 \includegraphics[width=0.49\textwidth]{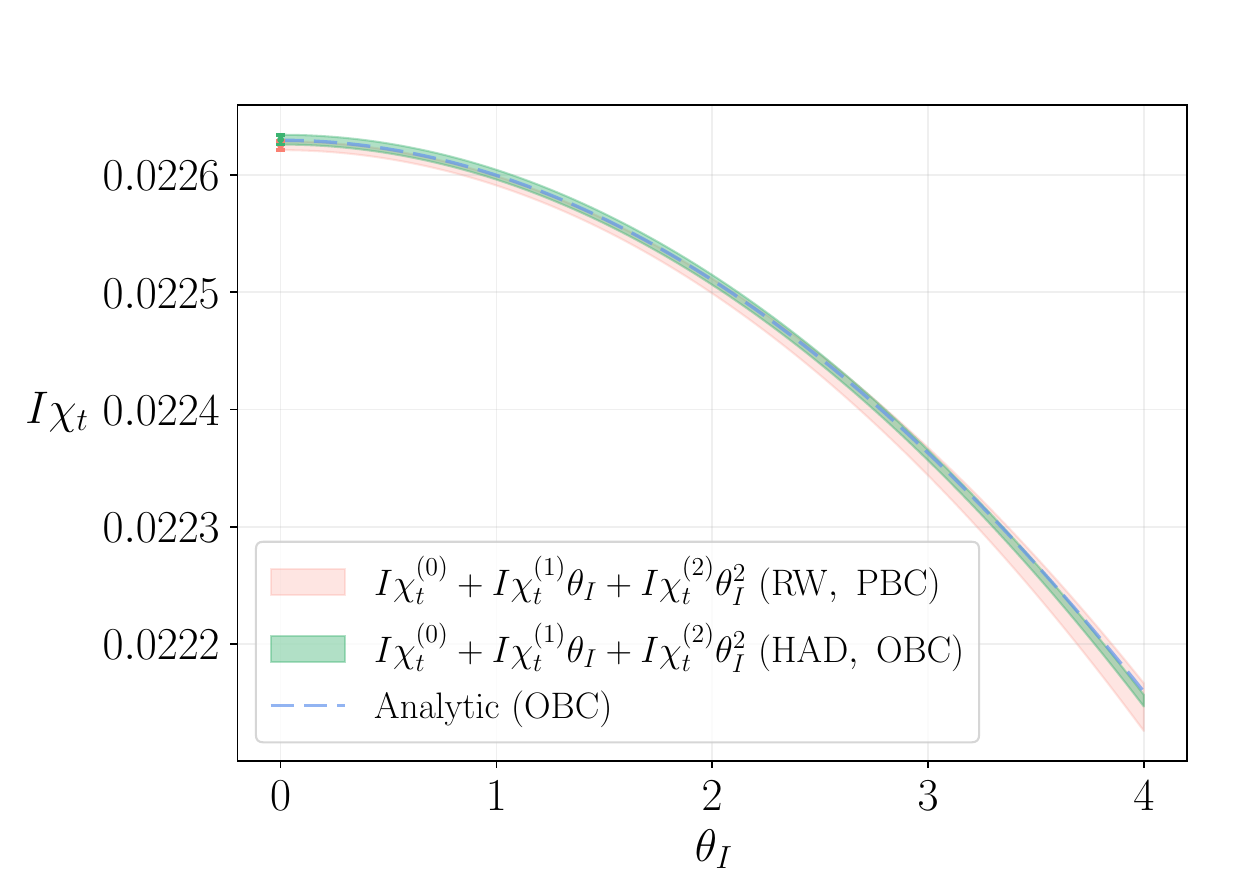}
 \caption{
   The $\theta_{I}$ dependence of $I \chi_{t}$ at $\hat{I}=5$ and $\hat{T}=100$
   computed with different methods for
   periodic boundary conditions and the standard definitions of the action and
   topological charge. (Left): results
   (orange circles) from five different simulations with 100k uncorrelated
   configurations each, along with their fit to the functional form
   $f(\theta_{I})=a+b\theta_{I} + c\theta_{I}^2$. (Right): results from single
   simulations with 500k uncorrelated configurations using reweighting and HAD
   (red and green points) at $\theta_{I}=0$, along with the curve
   $I\chi_{t}^{(0)}+I\chi_{t}^{(1)}\theta_{I}+I\chi_{t}^{(2)}\theta_{I}^{2}$ (light-red and thick-green
   bands) obtained from the use of truncated polynomials. The reweighting was
   performed on a simulation with periodic boundary condition, while HAD was
   used with open boundary conditions. In both panels the analytic result from
   open boundary conditions is displayed (dashed line).
 }
  \label{fig:theta-dependence-methods}
\end{figure*}

The sign problem is caused by the imaginary term in
Eq.~(\ref{eq:partition-function}), which for $\theta \neq 0$ is highly
oscillatory and leads to uncertainties which grow exponentially with the volume of the system. A conventional workaround
is to define $\theta_{I} \equiv i\theta \in \mathbb{R}$ so that the integrand becomes real and the
system can be simulated using standard sampling algorithms. By performing
simulations at different imaginary values of $\theta$, one can then use the
analiticity of an observable,
\begin{align}
O(\theta) = O^{(0)} + O^{(1)}\theta + O^{(2)} \theta^2 + \mathcal{O}(\theta^3), \quad O^{(n)} = \frac{1}{n!} \frac{\partial ^{n} O}{\partial \theta^{n}} \Big|_{\theta=0},
\end{align}
along with analytic continuation to obtain the expansion coefficients by
performing fits to data. This is depicted in
Fig.~\ref{fig:theta-dependence-methods} (left), we show results of $I \chi_{t}$ for different values of $\theta_I$.

An alternative method that allows to obtain arbitrarily high derivatives
of $\theta$ just from a single simulation at $\theta=0$ relies on the fact that a differentiable
function $f$ can be Taylor expanded around $x^{(0)}$.
The construction of such Taylor expansion can be automatized for any arbitrarily
complex function $f$ by using the algebra of truncated polynomials~\cite{haro_ad}:
if
$\tilde{x}=x^{(0)} + x^{(1)}\epsilon + x^{(2)}\epsilon^2 + \dots x^{(K)}\epsilon^{K}$
is a truncated polynomial of order $K$ and we code all elementary mathematical
functions of these polynomials, $f(\tilde{x})$ will be a truncated polynomial
containing the first $K$ derivatives of $f$.
Truncated polynomials are a particular automatic differentiation technique and
for our computations we use \texttt{FormalSeries.jl}~\cite{alberto_ramos_2023_7970278}, which can be
used for any arbitrarily complicated function $f$, such as a computer program implementing reweighting or the HMC algorithm.


A particularly simple application of truncated polynomials to extract higher
order derivatives from an existing simulation at $\theta=0$ is by
reweighting to $\theta\neq 0$ via the identity
\begin{align}
\langle O(\phi) \rangle_{\theta} = \frac{\langle e^{i\theta Q} O(\phi) \rangle_{\theta=0}}{\langle e^{i\theta Q} \rangle_{\theta=0}}.
\label{eq:rw-truncated}
\end{align}
By replacing $\theta$ with the truncated polynomial $\tilde\theta = \sum_{k=0}^{K}\tilde{\theta}^{(k)}\theta^{k}$ with
$\tilde{\theta}^{(1)}= 1$ and $\tilde{\theta}^{(k \neq 1)}= 0$, one automatically
obtains the full analytical dependence of the Taylor expansion
of  $\left<O(\phi) \right>_{\theta}$  with respect to $\theta$ up to order $K$
from a single ensemble. This is shown in Fig.~\ref{fig:theta-dependence-methods}
(right), where by reweighting a single standard simulation at $\theta_{I} = 0$
with periodic boundary conditions we automatically obtain $I\chi_{t}^{(k)}$;
particularly, we
see
that $I\chi_{t}^{(0)}+ I\chi_{t}^{(1)}\theta_{I}+ I\chi_{t}^{(2)}\theta_{I}^2$
agrees with the analytical result obtained with open boundary conditions.

\begin{table}[t]
    \centering
    \begin{tabular}{c|c|c}
        $\text{Method}$ & $\chi_{t}^{(0)} \times 10^{-3}$ & $\chi_{t}^{(2)}
        \times 10^{-6}$ \\
        \hline
      \hline
        Fit & 4.5238(16) & -6.08(48)\\
        Reweighting & 4.52501(76) & -5.99(25)\\
        HAD & 4.52604(83) & -5.980(34)\\
                          \hline
                          \hline
    \end{tabular}
    \caption{Comparison of errors between different methods to obtain the
$\theta_{I}$-dependence, namely, the quadratic fit to the results from direct
simulations at $\theta_{I}$ (Fig.~\ref{fig:theta-dependence-methods} left),
reweighting, and HAD (Fig.~\ref{fig:theta-dependence-methods} right). These results
correspond to simulations with $\hat{I}=5$ and $\hat{T}=100$, using the standard
definitions of the action and topological charge.}
    \label{tab:error-comparison-tab}
\end{table}

However, a drawback of reweighting is that the denominator in Eq.~(\ref{eq:rw-truncated}) has disconnected contributions
that are exponentially noisier with the size of the system. To solve this,
one can apply the truncated polynomials directly into the HMC
algorithm. Using the standard discretization of the model, the HMC equations of
motion read
\begin{align}
    \label{eq:eom-hmc-qr}
    \dot{\phi}_{t} &= \frac{\partial H[\pi, \phi]}{\partial \pi_{t}} = \pi_{t},  \\
    \dot{\pi}_{t} &= - \frac{\partial H[\pi,\phi]}{\partial \phi_{t}}
              = - I \left[ \sin(\phi_{t}-\phi_{t-1})- \sin(\phi_{t+1}-\phi_{t})
              \right]
               +\frac{\theta_{I}}{2\pi} \left[ \cos(\phi_{t}-\phi_{t-1}) -
               \cos(\phi_{t+1}-\phi_{t}) \right], \nonumber
\end{align}
where  $H[\pi,\phi] = \sum_{t=0}^{\hat{T}-1}\pi_{t}^2/2+ S_{\text{st}}[\phi] - \theta_{I} Q_{\text{st}}[\phi]$ is
the Hamiltonian of the system. By replacing $\theta_{I}$ by a truncated
polynomial, $\tilde{\theta}_{I}$, we obtain a Markov chain of $N$
samples, $\{\tilde\phi_{(i)}\}_{i=1}^{N}$, that carry the derivatives with
respect to $\theta_{I}$, and we denote this algorithm as Hamiltonian Automatic
Differentiation (HAD). The Taylor expansion of observables is obtained by
the computation of conventional expectation values using the
samples $\{ \tilde{\phi}_{(i)} \}_{i=1}^{N}$, and does not contain the noisy
disconnected contributions of the reweighting technique.\footnote{However, the Metropolis
accept-reject step is not differentiable and cannot be used with this method, so
one must integrate the equations of motion with high enough precision to avoid
systematic effects.} In Fig.~\ref{fig:theta-dependence-methods} (right), we show the curve
$I\chi_{t}^{(0)} + I\chi_{t}^{(1)}\theta_{I}+ I\chi_{t}^{(2)}\theta_{I}^2$
obtained from a single HAD simulation, and one can appreciate
that the predictions for high $\theta_I$ are more accurate than the ones
obtained by reweighting. The comparison can be seen more transparently in
Tab.~\ref{tab:error-comparison-tab}, where the error of the HAD algorithm is reduced by an order of magnitude with respect to the other methods at equivalent statistics.

\section{Results}
\label{sec:results}

\begin{figure}
  \centering
 \includegraphics[width=0.43\textwidth]{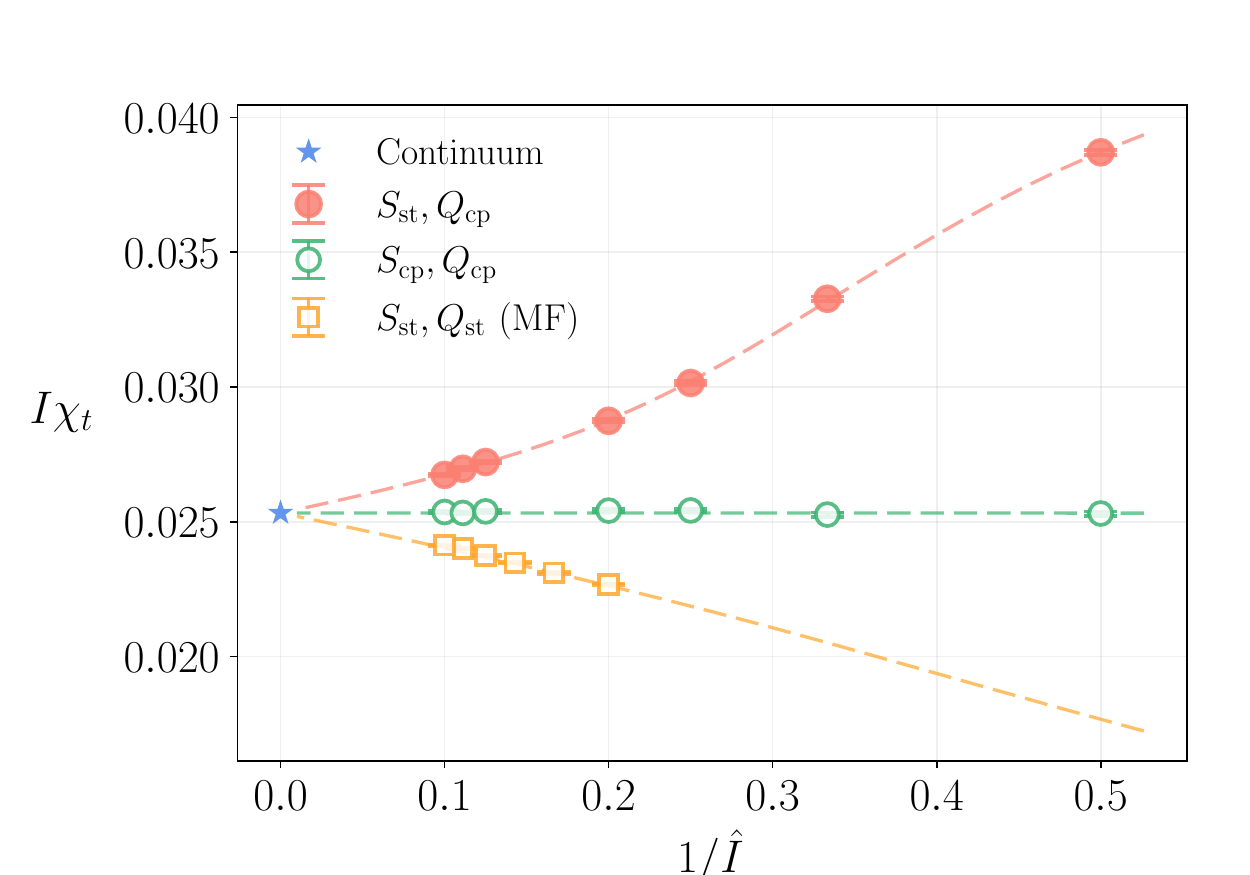}
 \includegraphics[width=0.43\textwidth]{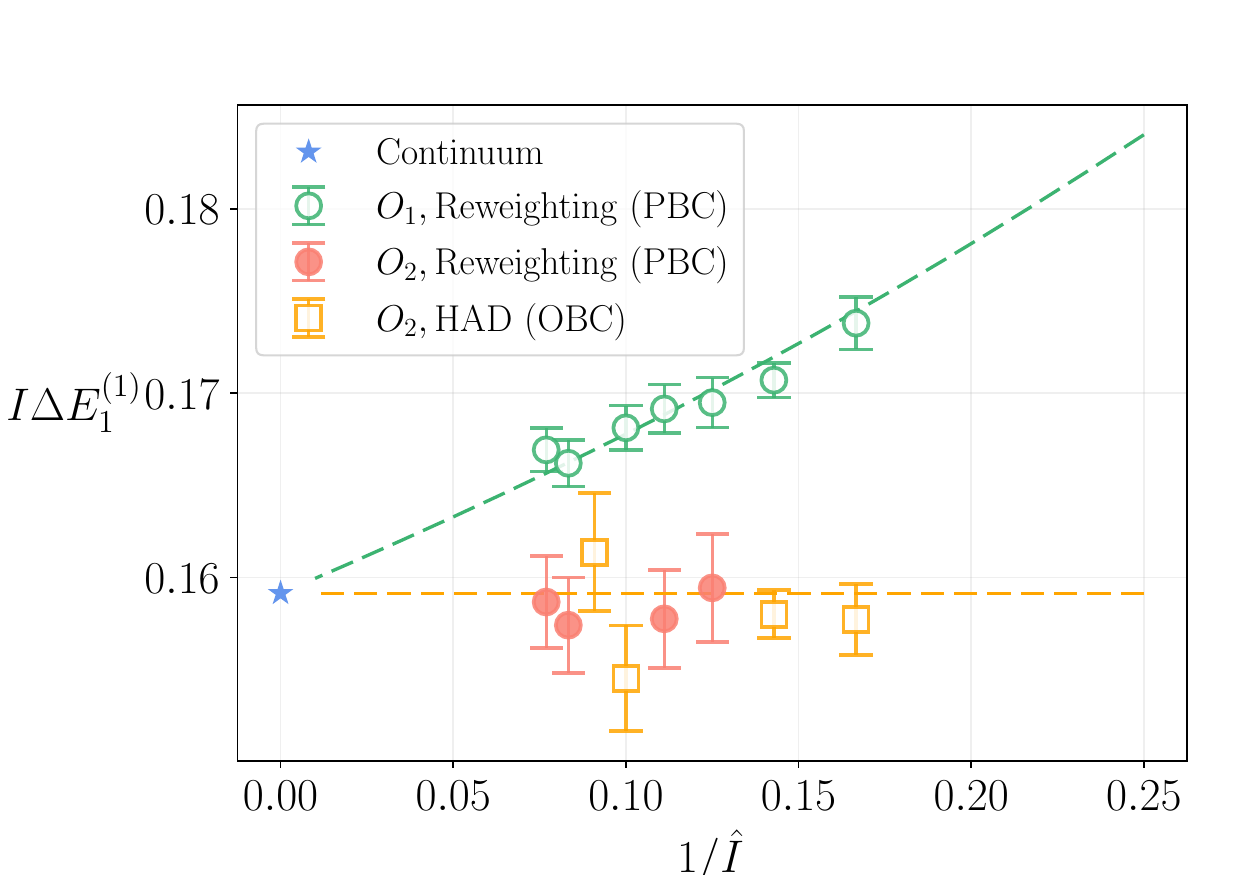}
  \caption{(Left) Continuum extrapolation of $\chi_t$ computed
from $ \langle (\phi_1 - \phi_0)^2 \rangle$ at various $\{\hat{T},\hat I\}$ at
constant $T/I=20$ with the standard action and classical topological charge
(red, filled circles), with the classical action and topological charge (green,
open circles) and from master field simulations with the standard action and
topological charge (orange, open squares), all with periodic boundary
conditions. The dashed lines represent the corresponding analytical results with
open boundary conditions. (Right) Linear $\theta$-correction to the first energy level, $\Delta E_{1}$, for a
    periodic lattice with the standard discretization of the action at different
    values of $\hat I$ with constant $T/I=20$. The results were obtained using
    the interpolating operator $O_{1}(t)=\phi_{t}$ with reweighting on a
    simulation with periodic boundary conditions (green, open circles), and
    using $O_{2}(t)=\sin(\phi_{t})$ with HAD (yellow, open squares) and
    reweighting (red, filled circles) with open and periodic boundary
    conditions, respectively.}
  \label{fig:chit}
\end{figure}

Fig.~\ref{fig:chit} (left) shows our results of a local version of the topological susceptibility, $\chi_t = \langle (\phi_1 - \phi_0)^2 \rangle$---with which the choice of boundary conditions and the quantization of the topological charge is irrelevant---from
simulations with periodic boundary conditions with both the standard and
classical perfect discretizations at different values of the lattice spacing.
The wHMC algorithm allowed us to perform simulations very close to the
continuum, and our results agree with the analytical results obtained with open
boundary conditions. All choices of boundary conditions and discretization lead to the continuum result in
Eq.~(\ref{eq:intro:qrotor-top-susc-E0}), validating the conventional order of
limits.

Finally, in Fig.~\ref{fig:chit} we show our results of the
linear $\theta$-dependence of the ground state of the spectrum for different
values of the lattice spacing, which from
Eq.~(\ref{eq:spectrum continuum})
reads
$\Delta E_{1} \equiv E_{1} - E_{0} = \frac{1}{2I} \left( 1 - \frac{\theta}{\pi} \right)$,
obtained through the usual spectral decomposition of an interpolator with the
same symmetries as the ground state. We use both reweighting and the HAD
algorithm, for different choices of boundary conditions and interpolating
operators, and conclude that the continuum limit agrees with the results
from quantum mechanics---and disagree with the claims of Refs.~\cite{ai_absence_2021,ai_consequences_2021}.

\section{Conclusions}

We have studied a recently proposed order of limits to study infinite-volume quantities which makes $\theta$
disappear from all physical observables of the theory. We have studied the continuum limit of the topological susceptibility and the first $\theta$-dependence correction to the ground energy state of the quantum rotor, validating the conventional wisdom on the strong $CP$ problem. Even though the system suffers from topology freezing and the sign problem, these were successfully overcome by the use of the wHMC algorithm and truncated polynomials. The generalization of these proposed algorithms to more complicated models is work in progress.

\section*{Acknowledgements}

We acknowledge support from the Generalitat Valenciana Grant
No.~PROMETEO/2019/083, the European Projects
No.~H2020-MSCA-ITN-2019//860881-HIDDeN and No. 101086085-ASYMMETRY, and the
National Project No.~PID2020-113644GB-I00, as well as the technical support
provided by the Instituto de Física Corpuscular, IFIC (CSIC-UV). D.A.
acknowledges support from the Generalitat Valenciana Grants No.~ACIF/2020/011
and No.~PROMETEO/2021/083. G.C. and A.R.  acknowledge financial support from the
Generalitat Valenciana Grant No.~CIDEGENT/2019/040.



\newpage

\bibliographystyle{JHEP}
\bibliography{references}

\end{document}